\documentclass[twocolumn,10pt]{tsfp}
\usepackage{flushend}
\usepackage{graphicx}
\usepackage[authoryear,round]{natbib}
\usepackage{fancyhdr}
\usepackage{amsmath, bm,amssymb}
\usepackage{url} 
\usepackage[hidelinks]{hyperref}
\usepackage{placeins}
\usepackage{xcolor}
\DeclareMathAlphabet{\pazocal}{OMS}{zplm}{m}{n}

\title{Sub-surface turbulence and free-surface features}

\author{Am\'{e}lie Ferran$^{1\dagger}$, Ali Semati$^1$, Ana\"{\i}s Rouaud$^{1,2}$, R. Jason Hearst$^1$, Simen {\AA}. Ellingsen$^1$
    \affiliation{
	$^1$Department of Energy and Process Engineering,
	Norwegian University of Science and Technology,\\
    7034 Trondheim, Norway \\
    $^2$Ecole Centrale M\'{e}diterran\'{e}e, Centre National de la Recherche Scientifique,
	Aix Marseille Universit\'{e},\\
	13384 Marseille, France\\
    $^\dagger$amelie.ferran@ntnu.no
    }	
 }

\begin{document}

\maketitle   
\thispagestyle{fancy}

\fontsize{9}{11}\selectfont

\section*{ABSTRACT}

Many turbulent flows encountered in nature---seas, oceans and rivers---are bounded by a deformable free surface.
A question that remained to be fully explored is to what extent the underlying turbulent flow field can be revealed solely by observing the surface deformations. 
In this study, we attempt to correlate free-surface topological deformations with the underlying turbulent flow field. 
We report an experimental investigation of the free surface in the wake of a surface-piercing cylinder and turbulence created by an active grid in an open-channel flow.
We are able to study instantaneous events of surface indentations and their related sub-surface coherent structures, as well as statistical properties of velocity and surface motion.
We observe weak cross-correlation between the vorticity field and the surface when considering the global surface elevation field.
Slightly stronger correlations emerge when conditioning the surface on specific regions, even in the case of three-dimensional homogeneous isotropic turbulence.

\section*{INTRODUCTION}

Near-surface turbulence governs important interactions between the atmosphere and the ocean, such as the exchange of gas, heat \citep{Dasaro2014,li2025}, the  mixing of biomatter \citep{Caldwell1995}, plankton \citep{Guasto2012} and pollutants \citep{DiBenedetto2025}. 
Small-scale surface deformations play a central role in these processes and are conventionally categorized into specific patterns, called `dimples’, `scars’, `ripples' and `boils’  \citep{brocchini_dynamics_2001}.
There are abundant observations that these surface features, such as scars and dimples, are directly related to turbulent flow structures \citep{banerjee94,longuet-higgins96,aarnes_vortex_2025}.
This naturally raises the question of how much of the underlying flow field can be revealed from the surface topology.

Previous studies have reported that point-to-point cross-correlations between flow field and surface quantities---such as vorticity and surface elevation---are weak when the surface is perturbed by homogeneous isotropic turbulence (HIT) \citep{savelsberg_turbulence_2008, savelsberg_experiments_2009}.
However, analysis conditioned on specific surface regions or patterns have shown a significant correlations between the occurrence of these features and certain flow-field-derived quantities, such as the horizontal divergence \citep{shen_surface_1999, dabiri_interaction_2003, babiker_vortex_2023, aarnes_vortex_2025}.
Experiments relating free-surface motion to sub-surface turbulence have mostly been statistical in nature \citep{dabiri_interaction_2003, savelsberg_experiments_2009}, although the potential to detect water-side flow features from their surface imprints has been explored \citep{Mandel2019, Gakhar2022} and in a field context \citep{muraro_freesurface_2021}.
There are few experimental works dedicated to identifying surface features and correlating them to the subsurface velocity field instantaneously; a recent effort in a zero-mean flow tank was performed by \citet{babiker2026}.

In this study, we measure the three-dimensional surface deformations and the underlying velocity field simultaneously. 
We explore two different approaches to correlate the interface topology with the turbulent flow beneath. 
In the current study we wish to compare our set-up to previous studies \citep{savelsberg_turbulence_2008, savelsberg_experiments_2009} and therefore regard point-to-point correlations between surface elevation and velocity field directly beneath. It was concluded by \citet{babiker2026} that point-to-point correlations rapidly fall off deeper than the thin near-surface viscous layer, approximately a quarter of the Taylor microscale thickness. All our measurements were performed below this depth.
First, we present point-to-point cross-correlations between surface elevation and vorticity fields for different incoming turbulent flows.
Second, we are able to identify and track instantaneous events of surface indentations and attempt to pair them with their corresponding subsurface coherent structures.

\section*{EXPERIMENTAL SETUP}

\begin{figure*}
	\centering
	\includegraphics[width=\textwidth]{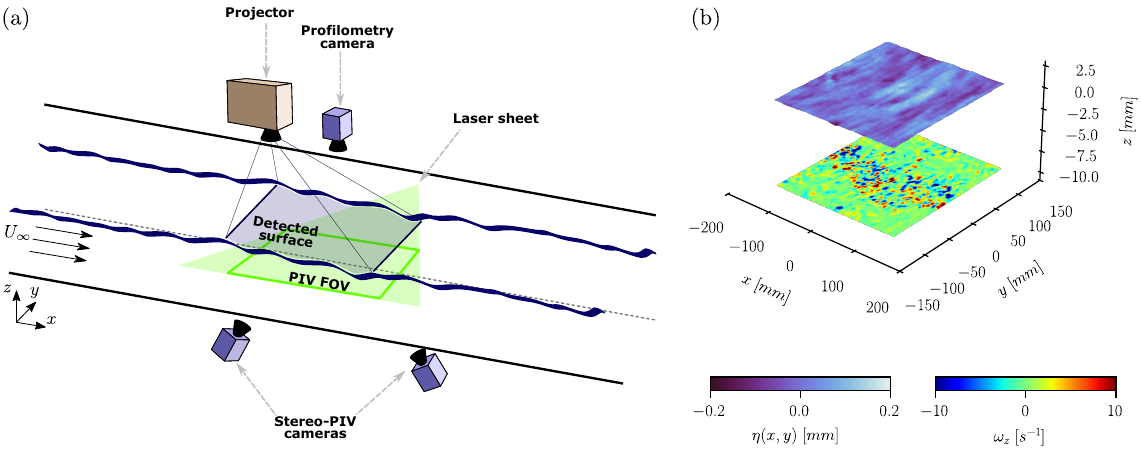}
	\caption{(a) Sketch of the SPIV/FPP experimental setup. The SPIV cameras beneath the water channel are configured to visualize the streamwise-horizontal plane, while the projector and the camera above the channel are used for the surface detection; the FPP field-of-view (FOV) is directly above that of the SPIV.
    (b) Example of an instantaneous horizontal out-of-plane component of the vorticity field, $\omega_z(x,y)$ and its corresponding surface elevation $\eta(x,y)$ around $30D$ downstream of the cylinder.}
	\label{figure_experimental_setup}
\end{figure*}

Free-surface turbulent flows generated by an active grid and a cylinder were investigated in the recirculating open-water channel at the Norwegian University of Science and Technology (Trondheim, Norway) \citep{jooss_spatial_2021}.
The subsurface velocity field and the three-dimensional surface topology were measured simultaneously using Stereoscopic Particle Image Velocimetry (SPIV) and Fringe Pattern Profilometry (FPP), respectively.


For the velocimetry measurements, two LaVision Imager CX2-25MP cameras, positioned beneath the water channel, captured an approximately $300$\,mm~$\times~300$\,mm horizontal plane 6\,m downstream of the test section inlet. 
The flow was illuminated with a double-pulsed Nd:YAG laser (Litron Nano L 200-15 PIV, wavelength $532\,\rm{nm}$) and seeded with $40\,\rm{\mu m}$ spherical polystyrene particles. 
Measurements were carried out in a horizontal plane parallel to the undisturbed water surface, at a depth of $z = 10$\,mm, with the water depth being $h=500$\,mm.


FPP is a surface measurement technique that consists of projecting a fringe pattern onto the water surface and capturing its deformations using a downward-pointing camera (here a LaVision Imager sCMOS) mounted above the channel.
A change in water surface height induces phase shifts in the projected image.
By comparing the pattern distortion with that of a reference image acquired on a flat horizontal surface, the three-dimensional surface topology can be reconstructed \citep{cobelli_global_2009}. 
To measure FPP and PIV simultaneously, the water surface must be made opaque to the projected light while remaining translucent to the PIV laser sheet. 
We achieved this by adding a fluorescent dye---fluorescein disodium salt hydrate (uranine)---at a concentration of 10~ppm, following the method recently introduced by \cite{semati2025}.

The two measurement systems were calibrated in tandem to ensure a common spatial origin, with the measured surface located directly above the PIV field of view (as illustrated in Figure~\ref{figure_experimental_setup}). 
The two acquisitions were synchronized, using different sampling rates of $45~\rm{Hz}$ and $15~\rm{Hz}$ for the FPP and the PIV, respectively.


Several inlet flow conditions were investigated: the planar wake of a surface-piercing cylinder, as well as homogeneous turbulence generated by an active grid.
A cylinder of diameter $D = 12$\,mm and length $400$\,mm was held above the water, piercing the water column up to a depth of $350$\,mm.
Planar wake measurements were performed between $20D$ and $40D$ downstream.
The active grid, based on the design of \cite{makita_realization_1991}, was situated at the entrance of the test section to generate three-dimensional 
HIT in the water bulk.
It consists of $18$ vertical and $10$ horizontal rods, four of the latter immersed in water, equipped with square-shaped wings each independently controlled by stepper motors. 
The grid mesh size $M$ is 100~mm, leading to the measurements being captured around $60M$ downstream the grid.
For this experiment, the grid vertical bars were randomly rotated at $0.05 \pm 0.025 ~ \mathrm{Hz}$. 
Two bulk velocities, $U = 0.25 ~\text{and}~ 0.38~\rm{m/s}$, were tested for both wake and grid-generated turbulence. 
Immediately downstream of the grid, a 1~m long plate lay flush with the surface to dampen surface-wave noise generated by the moving grid itself. 

For characterization of the turbulent flow, additional Laser Doppler Velocimetry (LDV) measurements were performed in the bulk.
Streamwise and vertical velocity components were acquired using a $60$\,mm FiberFlow probe (Dantec Dynamics) at a depth of $z=-200$\,mm.
From the LDV measurements, the flow integral length scale $L_\infty$ was obtained via integration of the streamwise velocity autocorrelation function until its first zero crossing. 
Assuming local isotropy and Taylor's frozen turbulence hypothesis, the Taylor microscale and the dissipation rate were estimated in the free stream from velocity gradients, using a 5-point centered difference scheme for the gradient calculation.
The Taylor microscale and the Taylor--Reynolds number were then defined as $\lambda^2 = (u')^2\langle (\partial u/\partial x)^2\rangle^{-1}$ and $Re_\lambda=u' \lambda / \nu$, with $u'$ the rms of the streamwise velocity fluctuations.
The different inlet conditions generated turbulent flows spanning a wide range of Taylor--Reynolds numbers as indicated in Table~\ref{tab:param}. 

\begin{table*}[ht]
    \centering
    \caption{Turbulent properties for the four experimental cases estimated using Laser Doppler Velocimetry at a depth of $z=-200$\,mm. For the wake cases, values correspond to the incoming background flow. 
    From left to right: mean bulk velocity $U$, rms of the streamwise velocity fluctuations $u'$, cylinder-based Reynolds number $Re_D = UD/\nu$, turbulent Reynolds number based on the integral scale $Re_T = 2 L_\infty u'/\nu$, Taylor-scale Reynolds number $Re_\lambda=u' \lambda / \nu$ and integral length scale $L_\infty$, with $\nu$ the water kinematic viscosity.}
    \label{tab:param}
    \begin{tabular}{ccccccccccc}
         & & & & & & & & \\ 
        \hline
        \hline
        & Grid & Cylinder & {\text{$ U ~ (m/s)$}} & {\text{$u' ~(m/s)$}} & {\text{$Re_D$}} & {\text{$Re_T$}}  & {\text{$Re_\lambda$}}  & {\text{$L_\infty  ~(m)$}} & {\text{$\lambda  ~(mm)$}} \\
        \hline
        C1 & Static & Y & 0.26 & 0.011 & 3120 & 370 & 85 & 0.016 & 7.8 \\
        C2 & Static & Y & 0.39 & 0.016 & 4680 & 574 & 159 & 0.018 & 10.6 \\
        A1 & Active & N & 0.25 & 0.023 & - & 11 224 & 379 & 0.244 & 17.6 \\
        A2 & Active & N & 0.38 & 0.037 & - & 22 274 & 680 & 0.301 & 19.4 \\
        \hline
        \hline
    \end{tabular} 
\end{table*}

\section*{RESULTS}
We first present some statistical properties of the surface deformations, then analyze the global cross-correlation between surface elevation and flow vorticity, and finally condition the flow field on the presence of surface features such as dimples and scars.

\subsection*{Surface measurements}

\begin{figure}[htb]
	\centering
	\includegraphics[width=0.88\columnwidth]{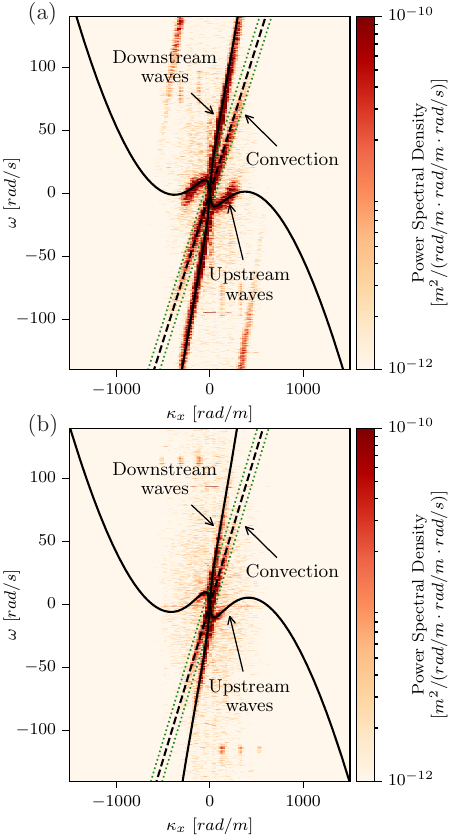}
	\caption{Power-spectral density of the surface elevation field $\eta(x,y)$ in the streamwise--temporal plane, i.e. the slice at $\kappa_y=0$, in the wake of the cylinder (a) and the active grid (b) for $U\approx0.25$\,m/s.
    Solid black lines indicate the wave dispersion relation; dashed lines correspond to $\omega=\kappa_x U$; dotted green lines indicate the Gaussian spectral filter bandwidth.}
	\label{figure_spectrum}
\end{figure}

We present in Figure~\ref{figure_spectrum} the wavenumber--frequency power spectral density of the free-surface motion for the cylinder case C1 (a) and the active grid case A1 (b) at the same incoming velocity $U\approx0.25$\,m/s. 
Here, the three-dimensional power spectrum is scaled by the frequency--wavenumber resolution to obtain the power-spectral density.
We consider the streamwise--temporal plane ($\kappa_x$, $\omega$), corresponding to the slice at $\kappa_y=0$, as it contains most of the relevant information.
The left half of the plane, $\kappa_x < 0$, is the redundant mirror image of the right half due to the complex representation of a real function.

Energy ridges of surface structures can be observed along the gravity-capillary wave dispersion relation, indicated by solid black lines in Figure~\ref{figure_spectrum}.
The dispersion relation of freely propagating waves on top of a uniform and steady current $\bm{U}\approx(U_x,U_y)$ ($U_x$ positive in the $x$-direction) is given by: 

\begin{equation}
    \omega_\pm(\bm{\kappa}) = \pm \sqrt{g\kappa +\sigma\kappa^3/\rho} + \bm{\kappa}\cdot \bm{U},
    \label{dispersion_relation}
\end{equation}
where $\sigma/\rho$ is the kinematic surface tension coefficient,  $\bm{\kappa}=(\kappa_x, \kappa_y)$ is the wave number vector and $\kappa = |\bm{\kappa}|$. Our coordinate system means $|U_y|\ll U_x$ and we write $\bm{\kappa}\cdot \bm{U}\approx\kappa_xU_x$.
The positive-signed (negative-signed) branch of the dispersion relation are the waves propagating in the downstream (upstream) direction in the reference system following the mean flow (the slowest upstream waves with phase velocity $\omega_+/\kappa_x\gtrless 0$ for $\kappa_x\gtrless0$, were swept downstream in the laboratory system).

Surface features passively advected by the mean flow velocity have spectral signal along the convection line, $\omega=\kappa_x U_x$, indicated as a dashed line in Figure~\ref{figure_spectrum}.
Imprints of turbulence on the surface---dimples, scars and upwelling boils---develop slowly (nearly `frozen') and the clear signal along the convection line in Figure \ref{figure_spectrum} is due to these. 
For both HIT and the planar wake, the surface signals from freely propagating waves and advected features have comparable strength, in contrast to the experiment by \cite{savelsberg_experiments_2009} which saw much stronger signal from waves; a likely explanation is the inlet surface plate which reduced the amount of wave noise significantly.
For the cylinder case, both contributions from waves and advected features are pronounced, due to the ``ship waves'' created by the surface-piercing cylinder and the coherent vortices shed by it.

\begin{figure}[tb]
	\centering
	\includegraphics[width=0.98\columnwidth]{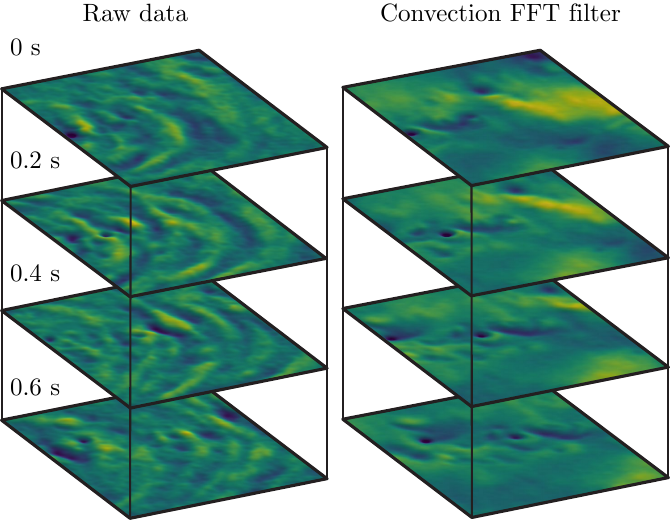}
	\caption{Surface images from FPP before and after applying the FFT convection filter.}
	\label{fig:FFTfilter}
\end{figure}

Surface deformations induced by waves decay exponentially with depth and for the wavelengths present in our experiment, give very weak velocity contributions already at $10$\,mm depth with correspondingly low correlations between surface motion and velocity field. 
This motivates restricting the correlation analysis to surface structures advected with the mean flow. 
To isolate these structures, the surface elevation fields were filtered in the frequency domain. The three-dimensional array of surface elevation, $\eta(x,y,t)$, was transformed to wavenumber--frequency space with a 3D discrete Fourier transform and a Gaussian mask was applied to retain spectral energy concentrated around the advective dispersion line $\omega=\kappa_xU_x$.
The filtered elevation field was then reconstructed by applying an inverse 3D FFT to the masked spectra and retaining the real part. The filter's spectral width was controlled by the Gaussian bandwidth, which was set to 15\,rad/s empirically, based on visual inspection of the filtered fields. Figure \ref{fig:FFTfilter} shows the effect of the filter on cylinder wake measurements for an example sequence where a particularly large turbulent feature passes by.

\subsection*{Global statistical correlations}

\begin{figure}[tp]
	\centering
	\includegraphics{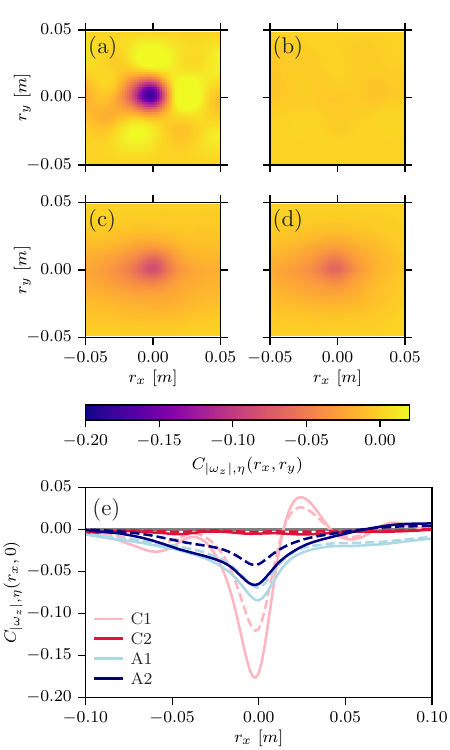}
	\caption{Cross-correlation between the surface shape and the vorticity field $C_{|\omega_z|, \eta}(r_x, r_y)$. 
    (a,b) Planar wake cases C1 (a) and C2 (b). (c,d) Fully developed HIT cases A1 (c) and A2 (d). 
    (e) Slices of the correlation function along the line $r_y=0$ for the four cases.
    Dashed lines denote correlation computed with the raw surface elevation field, while solid lines are used for the spectrally pre-filtered data.}
	\label{figure_global_correlations}
\end{figure}

We examine the cross-correlation $C_{|\omega_z|, \eta}(r)$ between the surface elevation $\eta(x, y, t)$ and the absolute value of the out-of-plane vorticity component $|\omega_z(x, y, t)|$, following \cite{savelsberg_experiments_2009}:

\begin{equation}
    C_{|\omega_z|, \eta}(r)=\frac{\langle |\omega_z(r+r', t)| \eta(r,t)\rangle_{r',t}}{\langle \omega_z^2 (r,t) \rangle^{1/2}\langle \eta^2 (r,t) \rangle^{1/2}},
    \label{correlation}
\end{equation}

where $\langle \cdot \rangle$ denotes time average and $\bm{r}=(r_x,r_y)$ is the horizontal separation vector with $r=|\bm{r}|$. 
The mean value was removed from each field before computing the cross-correlation.
Figure~\ref{figure_global_correlations}(a-d) presents the correlation maps $C_{|\omega_z|, \eta}(r_x, r_y)$ for the four flow cases, using the spectrally filtered surface elevation. 
The cross-correlation was computed over a spanwise band of approximately $[-2D, 2D]$, in order to restrict the analysis to the surface region located above the cylinder wake velocity deficit, where coherent vortices are present.
The same computational domain was used for the HIT cases.
Figure~\ref{figure_global_correlations}(e) presents the cross-correlation along the line $r_y=0$ for the four flow cases, where both the raw (dashed lines) and spectrally filtered (solid lines) surface elevation are considered.

It is expected that $|\omega_z|$ and $ \eta$ are negatively correlated due to the presence of surface-attached vortices associated with surface depressions, `dimples', corresponding to negative $\eta$ values. 
As expected, correlations are stronger in the low-speed planar-wake cases C1 than in the grid-turbulence cases, as the large coherent structures shed by the cylinder penetrate more deeply into the water column and dimples from shed vortices are clearly visible by eye.
For the high-speed wake case C2, the higher Reynolds number flow generates strong wave activity in the wake, making it difficult to detect any surface signature of turbulent eddies in the unfiltered spectrum, but pre-filtering the surface elevation to retain only advected flow structures improves the cross-correlation strength for all cases. 
The maximum absolute values of the cross correlation, after spectral filtering, are up to $\approx 0.18$ for the cylinder case and $\approx 0.08$ for the active-grid turbulence.

Although one is guaranteed to find a vortex directly beneath a dimple which typically extends beyond the near-surface viscous layer \citep{aarnes_vortex_2025}, dimples where the point-to-point correlation is strong make up a very small percentage of the overall surface, and the vortex tube will bend and no lie directly beneath the dimple, nor indeed be vertically oriented. 
Moreover, the thickness of the viscous layer within which point-to-point correlations can be strong when dimples and scars are first singled out \citep{aarnes_vortex_2025} is approximately $\lambda/4$ so our measurements are well outside of this layer in all cases.
The higher correlation values reported by \cite{savelsberg_experiments_2009} for vortex shedding behind a cylinder ($-C_{|\omega_z|, \eta}(0,0) \approx 0.4$) may be partly attributed to measurements performed much closer to the water surface (their PIV sheet depth was reportedly 1~mm beneath the surface, compared with 10~mm in the present study).
However, we report higher absolute correlation values for the HIT cases ($-C_{|\omega_z|, \eta}(0,0) \approx 0.04$ in \cite{savelsberg_experiments_2009, savelsberg_turbulence_2008}), possibly due to the presence of a plate at the test-section inlet used to damp surface-wave noise generated by the grid motion. 
The results of \citet{babiker2026} makes it likely that spatially non-local correlations between surface and bulk can be much higher, something we do not consider here.

\subsection*{Conditional correlations}

\begin{figure}[tp]
	\centering
	\includegraphics{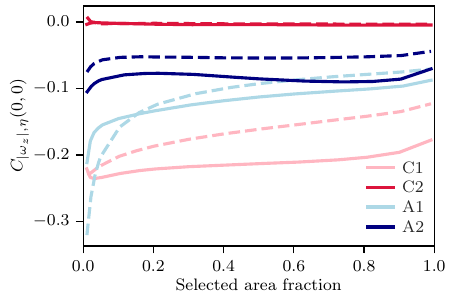}
	\caption{Conditional cross-correlation between surface elevation and vorticity at zero spatial separation, $C_{|\omega_z|, \eta}(0, 0)$, for the four flow cases, shown as a function of the fraction of the considered surface area. 
    Dashed lines denote correlations computed from the raw surface elevation, while solid lines correspond to spectrally pre-filtered data.}
	\label{figure_conditional_correlations}
\end{figure}

It has been shown by \cite{dabiri_interaction_2003} that conditioning free-surface elevations on negative-value regions is necessary to accurately estimate correlation coefficients in a vertical shear-layer flow.
Following this approach, we now condition our correlation analysis on surface depression regions.
To identify relevant regions with curvature characteristic of dimples and scars, we computed the wavelet transform of the free-surface elevation field.
Free-surface field wavelet transforms have previously been used to enhance the salient features \citep{babiker_vortex_2023, aarnes_vortex_2025}. 
A threshold criterion was then applied to the wavelet field to isolate regions associated with surface depressions.
A two-dimensional Mexican-hat wavelet, designed to highlight structures of about 10~mm, was employed as the mother wavelet.

Figure~\ref{figure_conditional_correlations} presents the conditional cross-correlation coefficient with zero spatial separation ($r_x=0, r_y=0$) as a function of the fraction of the domain considered.
Cross-correlation are computed with both spectrally filtered (solid lines) and raw (dashed lines) surface elevation fields. 
For most cases, the strength of the anti-correlation slowly increases as the considered area decreases.
When the area fraction is below 5\%, the selected surface regions consists mostly of dimples---associated with surface-attached vortices---or scars---associated with horizontal vortices located at the lower edge of the surface viscous layer. Since dimples now make up a much greater percentage of the area, a strong increase of the correlation absolute value is observed; however, while scars invariably have a vortex benath them, the corresponding vorticity vector is nearly horizontal with only a small $z$ component \citep{aarnes_vortex_2025}, so a strong correlation value is not expected even now, particularly for the cases without a cylinder present.

\subsection*{Surface features and flow vortices}
In this section, an attempt to identify and track instantaneous events of surface indentations and their corresponding sub-surface flow structures is described. 

Dimples and scars were detected using the computer-vision method of \cite{babiker_vortex_2023}, which relies on the fact that these surface features are either nearly circular or strongly elongated surface depressions, respectively, that persist over time.
Following \cite{babiker_vortex_2023}, the free-surface wavelet transform was thresholded to retain only 1\% of the total area.
The resulting events were tracked over time in the 45~Hz image time series and are required to satisfy a minimum lifetime of $0.2~\rm{s}$. 
Subsurface vortices were detected in the PIV field with a two-dimensional $\lambda_2$ criterion \citep{schram_wavelet_2004}. 
Similarly to the surface structures, vortices were tracked over time and only those persisting for at least 0.2~s (corresponding to three PIV frames) were retained.

\begin{figure}
	\centering
	\includegraphics{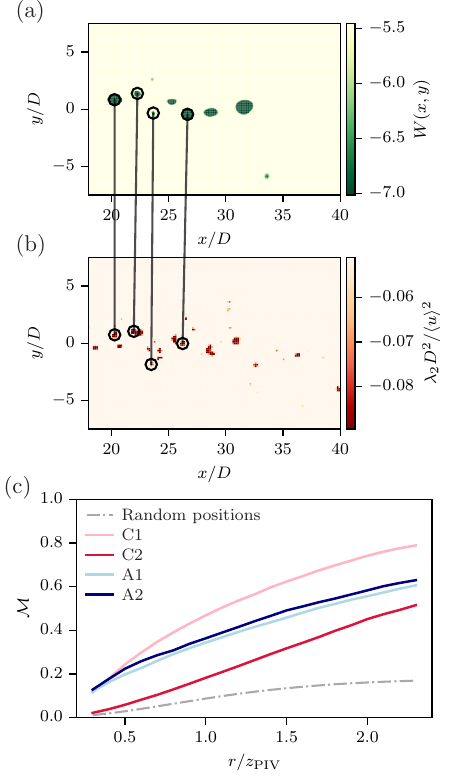}
	\caption{(a) Wavelet transform field of the surface elevation between $18D$ and $40D$ downstream of the cylinder.
    (b) Corresponding normalized $\lambda_2$ field.
    Matching events found with a search radius $r=z_{PIV}=10$\,mm that persisted over time highlighted with circles in both fields.
    (c) Matching ratio as function of the search radius $r$, normalized by the PIV sheet depth $z_{\rm{PIV}}$, for the different experimental cases.
    The probability of finding a vortex under a random point on the surface is shown with the gray dash-dotted line. 
    }
	\label{figure_dimples_vortices}
\end{figure}

Figure~\ref{figure_dimples_vortices}(a,b) shows the wavelet transform field $W(x,y)$ for a free-surface frame and the corresponding normalized $\lambda_2$ field downstream of the cylinder.
Green and red patches indicate regions below the respective selected thresholds, displaying surface depressions and high vorticity structures.
Dimple-vortex pairs were considered a match when the Euclidean distance between the two features was smaller than a search radius $r$. 
Persistent matches (evaluated with $r=z_\text{PIV}=10$\,mm) are highlighted with black circles in Figure~\ref{figure_dimples_vortices}(a,b).

We define \emph{match ratio}, $\pazocal{M}$, as the relative portion of detected dimples found to lie above a vortex.
Figure~\ref{figure_dimples_vortices}(c) shows $\pazocal{M}$ as a function of the search radius normalized by the average depth of the PIV plane $r/z_{\rm{PIV}}$, for the active grid and cylinder cases.
As expected, the match ratio increases with $r/z_{\rm{PIV}}$ as it introduces false positive matches.
When the search radius equals the PIV sheet depth, more than 25 \% of dimples coincide with an underlying vortex.
The probability of finding a vortex under an arbitrary surface point (grey dash-dotted line) is consistently far lower than under detected surface features.
Again, the dimple--vortex correspondence is weaker in the higher-velocity planar wake case (C2), most likely due to waves generated by the cylinder itself.

\section*{CONCLUSIONS and OUTLOOK}
The global cross-correlations between the surface elevation and the vorticity field beneath was found to give weak, but significant, maximum values when the correlation is calculated point-to-point. 
If the analysis is limited to a small percentage of the surface region containing persistent surface features, greater cross-correlation are found.
Finally, there is a larger probability of detecting a vortex under a detected surface feature, dimple or scars, than under an arbitrary position. The correlation values are weaker than those reported by \citet{savelsberg_experiments_2009} which we attribute mostly to their PIV measurements being much closer to the surface. 
It is known from studies of free-surface turbulence without mean flow that correlations between surface and bulk can be highly non-local \citep{babiker2026}, meaning that surface motion at one horizontal position can have a close connection with velocity features at the same time, but not directly beneath.
Future work will focus on correlating other quantities such as the horizontal flow divergence \citep{babiker_vortex_2023}, and explore non-local correlations.

\section*{ACKNOWLEDGEMENTS}
The work was co-funded by the European Research Council (ERC StG, GLITR, 101041000 and ERC CoG, WaTurSheD, 101045299) and the Research Council of Norway (iMOD, 325114). 
Views and opinions expressed are, however, those of the authors only and do not necessarily reflect those of the European Union or the European Research Council. Neither the European Union nor the granting authority can be held responsible for them.


\bibliographystyle{tsfp}
\bibliography{tsfp}


\end{document}